\documentclass[fleqn,12pt,twoside]{article}
\usepackage{espcrc1}

\usepackage{graphicx}
\usepackage[figuresright]{rotating}

\newcommand{\AmS}{{\protect\the\textfont2
  A\kern-.1667em\lower.5ex\hbox{M}\kern-.125emS}}

\title{Particle Correlations at RHIC -- Scrutiny of a Puzzle}

\author{Sven Soff$\,$\address{Lawrence Berkeley National Laboratory, 
Nuclear Science Division MS70-319,\\ 
1 Cyclotron Road, Berkeley, CA94720, USA},
Steffen A.\ Bass\address{
       Department of Physics, Duke University, Durham, NC27708, USA, and\\
       RIKEN BNL Research Center, Brookhaven National Laboratory, 
       Upton, NY11973, USA},
David H.\ Hardtke$^{\rm a}$
        and
Sergey~Y.~Panitkin\address{
Physics Department, Brookhaven National Laboratory, 
PO Box 5000, Upton, NY11973, USA}
}

\begin{document}

\maketitle
\vspace{-0.3cm}
\begin{abstract}
   We present calculations of two-pion and two-kaon  
   correlation functions in relativistic heavy ion collisions 
   from a relativistic transport model that includes explicitly  
   a first-order phase transition from a thermalized quark-gluon plasma 
   to a hadron gas. We compare the obtained correlation radii 
   with recent data from RHIC. The predicted 
   $R_{\rm side}$ radii agree with data while the $R_{\rm out}$ and 
   $R_{\rm long}$ radii are overestimated. 
   We also address the impact of in-medium modifications, for example, 
   a broadening of the $\rho$-meson, on the correlation radii. 
   In particular, the 
   longitudinal correlation radius $R_{\rm long}$ is reduced, 
   improving the comparison to data.   
\end{abstract}
\vspace{0.4cm}

Two-particle correlations at small relative momenta 
have been predicted to be particularly 
sensitive to a phase transition from quark-gluon matter 
to hadronic matter \cite{pratt86}. 
For a first-order phase transition, larger 
hadronization times were expected to lead to considerably enhanced 
correlation radii, 
characterizing the space-time extension of the particle-emitting source, 
compared to, for example, a purely hadronic scenario. 
The radii should also depend on 
the critical temperature $T_c$, the latent heat, 
the initial specific entropy 
density or the initial thermalization time of the quark-gluon phase.  

Here, we discuss relativistic transport calculations at RHIC energies 
that describe the initial dense  stage 
by hydrodynamics \cite{DumRi} and the later more 
dilute stages by microscopic transport \cite{bass98} of the particles.
The two models are matched at the hadronization hypersurface \cite{hu_main}. 
In the hadronic phase the particles are allowed to 
rescatter and to excite resonances 
based on cross sections as measured 
in vacuum. 
For the initial dense (hydrodynamical) phase a bag 
model equation of state 
exhibiting a first-order phase transition is employed. 
Hence, a phase transition in local equilibrium that proceeds 
through the formation of a mixed phase, 
is considered. The details of this relativistic hybrid transport model 
can be found elsewhere \cite{hu_main}. 

We first briefly summarize the main conclusions obtained 
in previous work  \cite{hirschegg,kaonlett,soffbassdumi}. 
Then, we show in detail the results for pions
when calculating explicitly the correlation functions from 
the source function of the transport model. 
These results are subsequently compared to calculations that take in-medium modifications 
into account and to experimental data for central Au+Au collisions at 
$\sqrt{s}_{NN}=130\,$GeV 
\cite{STARpreprint,Johnson:2001zi}.

Model calculations have demonstrated that the 
freeze-out hypersurfaces of pions extend to rather large radii and 
times compared to the  size of the mixed phase \cite{soffbassdumi,hirschegg}.  
In this hadronic phase many soft collisions take place 
which hardly modify the single-particle spectra 
but have a strong impact on the 
correlation functions that measure the final freeze-out state. 
Studying this in detail leads to the following conclusions \cite{soffbassdumi}: 
(i)  The dissipative hadronic phase leads to a rather large duration of emission.
(ii) The $R_{\rm out}/R_{\rm side}$ ratio, thought to be a characteristic measure of this 
emission duration, increases with transverse momentum.
(iii) The specific dependencies of the interferometry radii on the QGP properties are rather weak 
due to the dominance of the hadronic phase. This even results in qualitative 
differences if calculations with and without this subsequent 
hadronic phase are compared (dependence on the critical 
temperature) \cite{soffbassdumi}.

The correlations of kaons provide a severe test of the 
pion data and have several 
advantages \cite{kaonlett}. 
In particular, the kaon density is much lower than the pion 
density \cite{Murray:2002ek}. 
Hence, multiparticle correlations that might play a role for the pions are 
of minor importance for the kaons. Also, 
the contributions from long-lived resonances are under 
better control for kaons. 
The $R_{\rm out}/R_{\rm side}$ ratio for kaons is shown in Fig.~1. 
The sensitivity to $T_c$ and to the specific entropy density 
(SPS vs.\ RHIC) is enhanced 
at larger transverse momenta ($K_T \sim 1\,$GeV/c). 
This enhanced sensitivity is also driven by a strong increase of 
the direct emission component 
(from the phase boundary) at high $K_T$ as shown in Fig.~2. 
More and more kaons (up to $\sim 30\%$) escape  
the initial stages (unperturbed by the hadronic phase).\vspace*{-0.8cm} 
\begin{figure}[htb]
\begin{minipage}[t]{76mm}
\includegraphics[width=75mm]{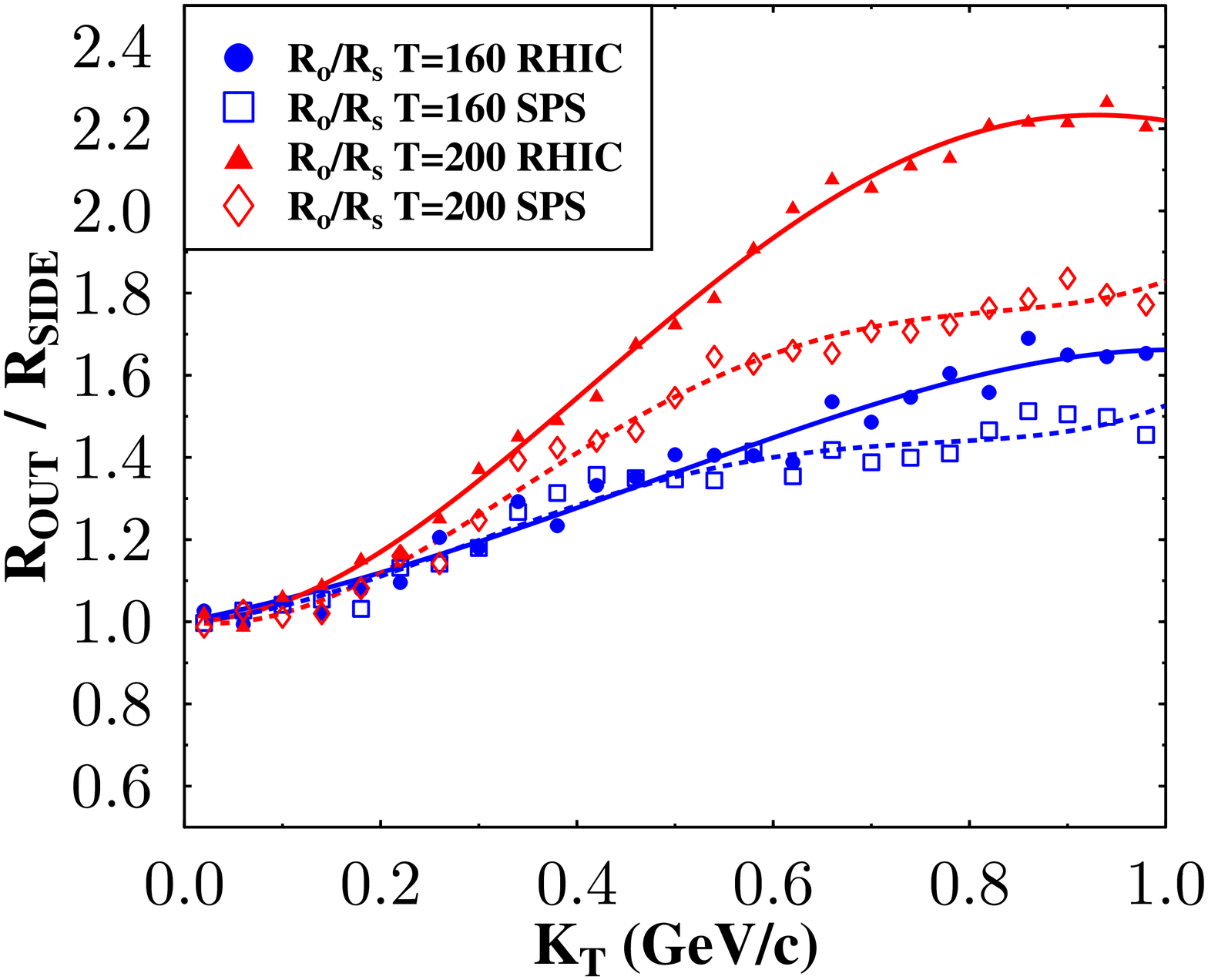}
\mbox{ }\vspace{-1cm}
\caption{$R_{\rm out}/R_{\rm side}$ for kaons at RHIC (full symbols) and
at SPS (open symbols), as a function of $K_T$ for critical temperatures
$T_c\simeq 160\,$MeV  and $T_c\simeq 200\,$MeV, respectively.}
\label{fig:largenenough}
\end{minipage}
\hspace{\fill}
\begin{minipage}[t]{79mm}
\includegraphics[width=78mm]{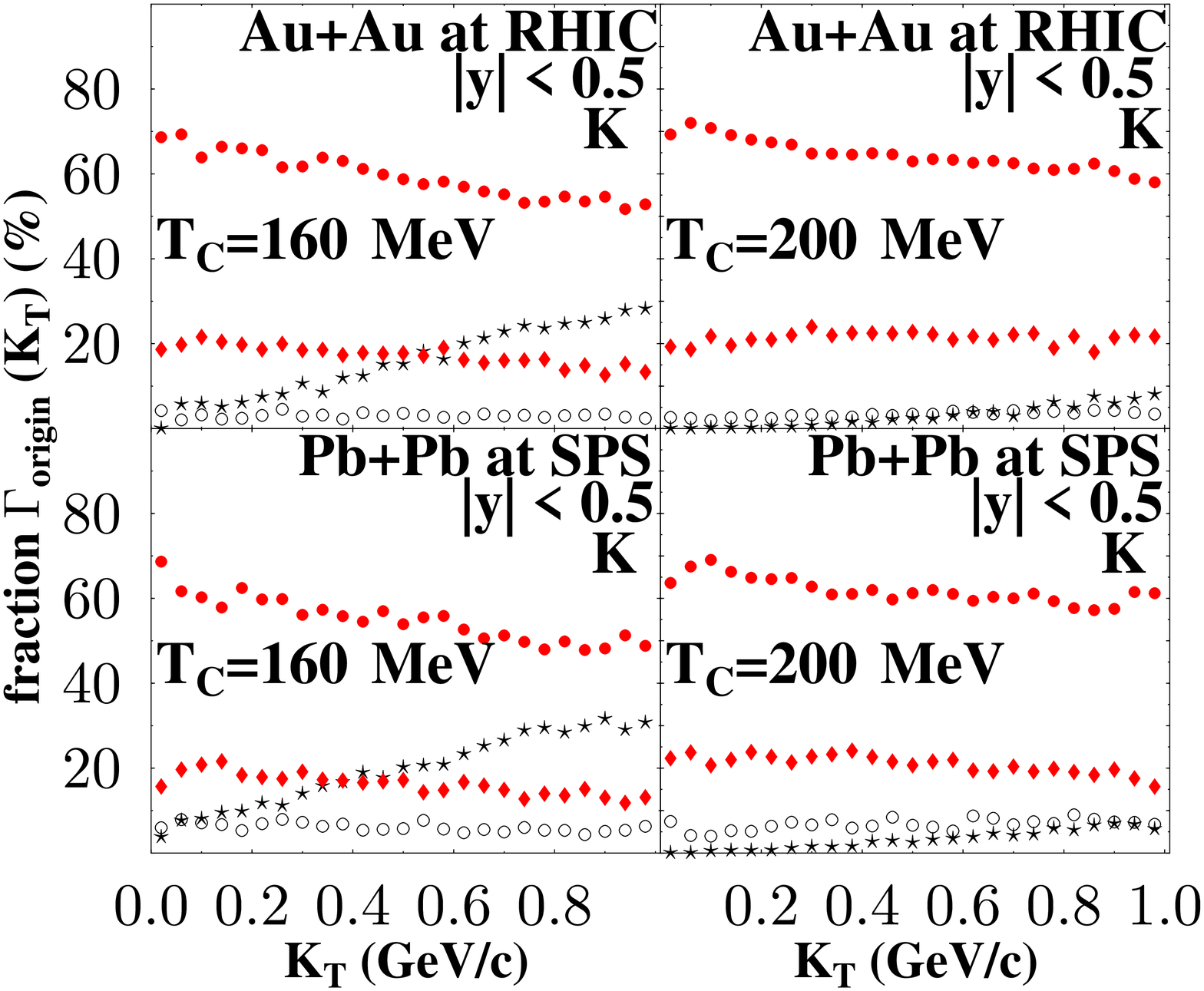}
\mbox{ }\vspace{-1cm}
\caption{Fraction of kaons $\Gamma_{\rm origin}$ that origin from a particular
reaction channel prior to freeze-out. These are resonance
decays (full circles),  direct emission from the phase boundary (stars),
elastic meson-meson (diamonds), or elastic meson-baryon (open circles)  
collisions.}
\label{fig:toosmall}
\end{minipage}
\end{figure}

The pion correlation parameters 
$R_{\rm o},\,R_{\rm s},\,R_{\rm l}\,$and $\, \lambda$ 
are obtained from the explicit calculation of the correlation functions 
\cite{Pratt:1990zq} in the 
respective 
transverse momentum bins and subsequent fitting of these 
correlation functions to 
a Gaussian form of the correlator $C_2=1+\lambda \exp(-R_{\rm o}^2q_{\rm o}^2
-R_{\rm s}^2q_{\rm s}^2-R_{\rm l}^2q_{\rm l}^2)$. 
The results of fitting the 3-dimensional pion 
correlation functions (as obtained from the transport calculations 
(RHIC initial conditions, $T_c\approx 160\,$MeV) + {\it correlation
after burner} (by Pratt) \cite{Pratt:1990zq})  
are shown, together with 
the experimental STAR and PHENIX data  
\cite{STARpreprint,Johnson:2001zi}, in Figure~3.
Even for a first-order phase transition scenario,  
the calculated interferometry radii (T) 
are not unusually large. 
Finite momentum 
resolution (f.m.r.) reduces the {\it true} correlation radii 
and the $\lambda$ intercept parameter, 
in particular, at higher $K_T$ \cite{kaonlett}.
This is corrected for in the experimental analysis. 
\begin{figure}
\begin{minipage}[t]{120mm}
\includegraphics[width=119mm]{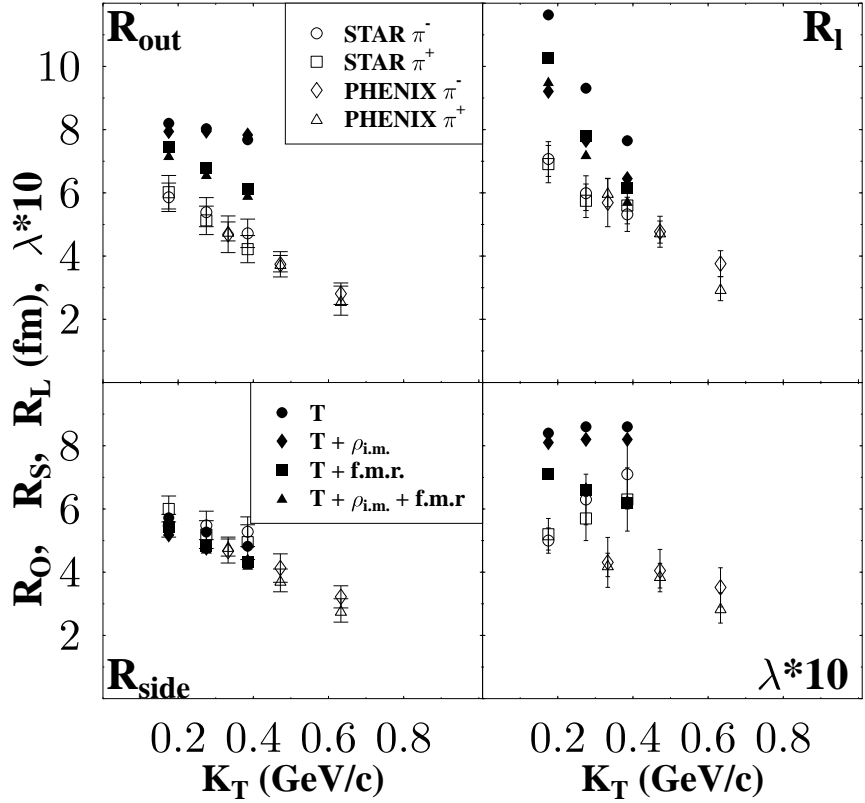}\\
\end{minipage}
\vspace{-1.0cm}
\caption{Pion correlation parameters $R_{\rm o},\, R_{\rm s},\, 
R_{\rm l},\,$and$\,\lambda$ as obtained from 
the `QGP+hadronic rescattering' transport model 
calculations + {\it correlation after burner} (T) compared to 
experimental data from PHENIX and STAR. For theory (T), the effects 
of a finite momentum resolution (f.m.r.) and a broadened in-medium 
$\rho$ spectral function ($\rho_{\,\rm i.m.}$) are calculated. 
Note the strong reduction of $R_l$ when an in-medium $\rho$ 
is considered.\vspace{-1.1cm}}
\label{pionradii}
\end{figure}
While the $R_{\rm side}$ radii show agreement  
the $R_{\rm out}$ radii 
are too large compared to the data.
However, these pion radii are considerably smaller 
than the corresponding radii obtained 
from the coordinate-space points and using expressions for the 
Gaussian radius parameters based 
on a saddle-point integration over 
the source function \cite{hirschegg,kaonlett,soffbassdumi,Hardtke:2000vf}.  
Only the calculations as shown in Figure~3 which are obtained 
from the complete calculation 
and the performed fits should be compared to data. 
The $R_{\rm out}/R_{\rm side}$ ratio is also larger
than unity for the fitted values and 
confirms the so-called {\it HBT-puzzle}, i.e., 
the RHIC data from STAR and PHENIX \cite{STARpreprint,Johnson:2001zi} 
indicate a decreasing ratio with $K_T$ \mbox{(even below 1)} 
and the calculations show only ratios larger 1.  
We remark that at the CERN-SPS (see Refs.\ in \cite{hirschegg})
the $R_{\rm out}/R_{\rm side}$ ratio seems to increase 
with $K_T$, being larger than 1, implying  
a real qualitative change of the reaction 
dynamics from SPS to RHIC energies.
A discussion of theoretical assumptions and uncertainties, experimental 
corrections, and possible solutions of this problem 
(including strongly opaque sources or scenarios with strong supercooling
leading to 
spinodal instabilities) is provided in \cite{hirschegg}.  

In-medium modifications may alter the correlation radii due to their 
impact on the mean free path via modified cross 
sections at finite temparatures or densities 
($T_{\rm f.o.}\sim 100-120\,$MeV). 
Increasing the width of the $\rho$ from the vacuum value of 
$\Gamma_{\rho}=150\,$MeV to $\Gamma_{\rho}^{\rm i.m.}=350\,$MeV 
enhances the opacity of the system (more $\pi \pi \leftrightarrow \rho$ 
processes), with a particular reduction 
of the radius $R_{\rm long}$, improving the 
comparison to data (Fig.~3). 

We have demonstrated that the kaon interferometry measurements, in particular 
at high $K_T$,  
will provide an exellent probe of the space-time dynamics 
(close to the phase boundary). 
In addition they represent a severe test of the pion correlations 
and may help to better understand the {\it HBT-puzzle} 
\cite{Zschiesche:2001dx}, i.e., the 
difference in the $K_T$ dependence of the $R_{\rm out}/R_{\rm side}$ ratio 
between the model predictions and the experimental RHIC data. 
The differences are due to the  
$R_{\rm out}$ radii  (which are larger in the model calculations) 
while the $R_{\rm side}$ radii seem to be described reasonably. 
An in-medium broadening of the $\rho$-meson reduces in particular 
$R_{\rm long}$, improving the comparison to data.

\noindent{\bf Acknowledgments: }
SS is supported by the Alexander von Humboldt Foundation through a 
Feodor Lynen Fellowship and DOE Grant No.\ DE-AC03-76SF00098.

\end{document}